\RequirePackage{fix-cm}
\documentclass[smallcondensed]{svjour3}     
\smartqed  
\usepackage{graphicx}
\usepackage{amsmath}
\usepackage{amsfonts}
\newcommand{\Z}{\ensuremath{\mathbb{Z}}}
\newcommand{\R}{\ensuremath{\mathbb{R}}}
\newcommand{\M}{\ensuremath{\mathcal{M}}}
\newcommand{\Mo}{M\"obius}

\journalname{International Journal of Theoretical Physics}
\begin{document}

\title{Electrodynamics and time orientability}


\author{Dr Mark J Hadley}


\institute{\at
              Department of Physics, University of Warwick, Coventry
CV4~7AL, UK \\
              Tel.: +44 77 2406 2496\\
              \email{Mark.Hadley@warwick.ac.uk}           
}

\date{Received: 5 January 2017 / Accepted: 6 March 2017}

\maketitle

\begin{abstract}
On spacetimes that are not time orientable we construct a U(1) bundle to measure the twisting of the time axis. This single assumption, and simple construction, gives rise to Maxwell's equations of electromagnetism, the Lorentz force law and the Einstein-Maxwell equations for electromagnetism coupled to General relativity. The derivations follow the Kaluza Klein theory, but with the constraints required for connections on a U(1) bundle rather than five spacetime dimensions. The non time orientability is seen to justify and constrain Kaluza Klein theories exactly as required to unify gravitation with electromagnetism. Unlike any other schemes, apparent net electric charges arise naturally because the direction of the electric field reverses along a time reversing path. The boundary of a time reversing region can therefore have a net electric flux and appear exactly as a region containing an electric charge. The treatment is purely classical, but motivated by links between acausal structures and quantum theory.


\end{abstract}
\section{Introduction}
Given a U(1) bundle over a 4D spacetime manifold, it is well known that the curvature of the bundle gives the same equations as the source-free Maxwell equations (The derivation is given in a many lecture notes and books on vector bundles and differential geometry \cite{eschrig2011topology,MTW,naber2011topology} for example). Here we consider some of the consequences of spacetime not being time orientable, and use the non-orientability of time to define a U(1) bundle over a spacetime manifold.  Our construction not only gives a geometric origin to the U(1) bundle, it also goes further in deriving the equations of electromagnetism including some aspects of quantisation. Although this paper is entirely classical, the lack of a time orientation gives new results and is motivated by earlier work linking spacetimes that are not time orientable with foundations of quantum theory \cite{hadley97,hadley98}, the existence of electric charges \cite{diemer_hadley}, and spin half transformations of elementary particles \cite{hadley2000}.

Orientability is a global property, locally every manifold is orientable. It is a global topological property that is not directly related to intrinsic curvature. A plane, cylinder and \Mo\ band are all flat manifolds, yet only the \Mo\ is non orientable. It should therefore be clear that the intrinsic curvature and hence Einstein's equations of general relativity don't give information on the orientability. The manifold can twist and turn independently of the intrinsic curvature. In differential geometry, information about the "twisting and turning" of the tangent plane is held in the connection one-forms. In this paper we add the simplest additional structure to spacetime that allows us to use smooth functions on the manifold to characterise the time orientability. .

If  a spacetime is not time-orientable then a closed path exists round which the direction of time reverses. The simplest example of non-orientability is the \Mo\ strip. On the \Mo\ strip left-handed and right-handed cannot be consistently defined over the whole surface. A left-handed coordinate basis changes to a night-handed one when going round the circumference of the strip. The \Mo\ band can also be thought of as a spacetime diagram for a circular space, $S^1$, and a non-orientable time. The direction of time reverses on a path around the circumference $S^1$ of the band. Note that our usual image of a \Mo\ strip is as a 2D surface embedded in 3D. However the embedding is not unique and the \Mo\ can be defined in a number of ways without resorting to any embedding at all. More importantly, it has topological properties (The non orientability) that can be described independently of the embedding.

Of particular interest is a model of a particle as an asymptotically flat spacetime manifold with a region of non trivial topology where time is not orientable. We therefore have a classical model for a particle and structures which can be used to derive equations similar to electromagnetism. As far as possible the treatment and results will be generic, independent of any precise spacetime structure. But as an example one could consider a wormhole structure where time reverses when passing through it. The famous Einstein Rosen bridge is such an example\cite{einstein_rosen}, although modifications to make it traversable are speculative. Mathematically two balls (worldtubes) are removed from $\R^3$ ($\R^3 \times \R$) and their surfaces are identified to create the wormhole. it is non time orientability if the identification reverses the time direction.

Historically, there have been other approaches to add the extra structure to spacetime to derive some of Maxwell's equations. Two notable cases are Kaluza and Klein adding a 5th dimension to gravitation (Wikipedia has a remarkably good introduction: 
) and then, following the approach of general relativity, to derive combined equations for gravitation and electromagnetism. It is an obvious extension to general relativity, somewhat undermined by important arbitrary restrictions on the 5th dimension. It succeeds in deriving the lorentz force law, but fails to describe charges. By rolling up the 5th dimension into a small compact co-ordinate, it effectively creates a U(1) bundle, but uses an assumed metric to derive equations. A second commonly described motivation for the U(1) structure is to take the phase of the complex quantum mechanical wavefunction. Source free Maxwell equations can be derived and some structures are known where the topology can induce a quantisation condition for charge. It is the first and simplest Yang Mills theory. This paper gives a very different physical meaning to the additional U(1) structure. It is purely classical and the results have a clear geometric origin.

\section{The mathematical structure}
Let $\M $ be a spacetime manifold endowed with a Lorentzian metric; we further require $\M $ to be asymptotically flat. We model a particle as a region of $\M $ with non-trivial topology that is not time orientable. We can remove the worldtube, $T_w$ of the particle leaving an asymptotically flat region $U_\infty = \M - T_w \simeq R^{3,1} - (B^3 \times \R )$ which has an approximately Minkowski metric. We make no assumptions about the region $T_w$ except that it is not time orientable and $\partial T_w \simeq \partial (B^3 \times \R)$.  $U_\infty$ is crucial, it is the spacetime that we experience and do our experiments in. The whole manifold $\M $ is not time orientable, but $U_\infty$ is both space and time orientable; it is the consequences of the lack of a time orientation that are studied in this paper. The exciting aspect of this work is to understand how the non-trivial part of the manifold has observable effects in region $U_\infty$.

A time orientable manifold can have a Lorentzian metric if and only if there exists a global timelike unit vector field (see for example O'neill\cite[p145]{oneill}). For a manifold that is not time orientable a vector field exists up to sign. In other words a time axis is well defined, even though the positive and negative directions cannot be defined globally. (If that is not obvious, a vector field can be constructed on the orientable double cover and then projected onto $\M$). Of course the vector field is not unique, but the metric structure embodies special relativity and ensures that the timelike vector field is timelike and unit length for all observers. In the asymptotically flat region we can choose a normal timelike vector $\hat{t}_\infty$ which is unique up to Lorentz transformations and sign. The following constructions do not require a unique timelike vector field.

On an orientable spacetime, on any closed loop, the timelike vector field can be consistently defined:one might imagine the unit normal vector twisting or turning, but the result of any \emph{rotation} is a multiple of $2\pi$. On a non time orientable manifold, some paths will result in $\hat{t}$ changing to $-\hat{t}$. This can be treated as a smooth change along the path by adding a notional angle $\theta$ which must integrate to a multiple of $\pi$ on any closed curve. This structure is a U(1) bundle. On a non time orientable manifold the bundle is necessarily non-trivial.

Formally: on any patches, A,B, of the spacetime manifold we have a trivial U(1) bundle. At the overlap of the patches the fibres are related by:
\begin{equation}
  \begin{array}{lllll}
    \theta_B &=& \theta_A & \textrm{if} & \hat{t}_B=\hat{t}_A  \\
    \theta_B &=& \theta_A +\pi & \textrm{if} & \hat{t}_B=-\hat{t}_A  \\
  \end{array}
\end{equation}

This is the same structure as the orientation bundle. There are a number of different ways to regard the construction. It could be seen as an extra dimension for the time direction to rotate in, which is close to the original Kaluza Klein hypothesis. Formally we add an extra timelike dimension, locally $M \otimes \R$. A unit timelike vector $\hat{t}$ has an associated angle $\theta$ which measures where it points in the time plane. However the full freedom of an extra spacetime direction is not required and leads to extraneous fields and parameters that cause problems of interpretation in the Kaluza Klein theory. A closer analogy is with an embedding of M into one higher dimension. Alternatively it can be regarded as a complex time co-ordinate with the U(1) parameter being the complex phase, but this implies unnecessary analytical complications when taking time derivatives. Note that the orientable double cover of the manifold is a subset of the U(1) bundle with $\theta = 0, \pi$ corresponding to the two orientations on the double cover. The way $\theta$ changes around the manifold is determined by the connection one-form, which in a sense captures the twisting of the time direction.

It is worth noting the very significant difference between the U(1) bundle described here and the bundles formed from the complex phase of a quantum mechanical wavefunction, which is the common interpretation and construction of a U(1) bundle associated with electromagnetism. The approach in this paper is purely classical and any consequential quantisation is classical in origin and not an artifact of starting with a quantum theory.

\subsection{Gauge transformations and embeddings}
The choice of $\theta = \theta(p)\ p \in \M$ on any patch of the manifold is largely arbitrary. A point by point change to $\theta$ is simply a gauge transformation. Our construction made use of a global time axis, which exists but is not unique. Different timelike axes are related by Lorentz transformations which change the coordinate system on a patch of the manifold and \emph{rotate} the time direction within the tangent bundle, this has no direct effect on the U(1) bundle construction.

\section{Electromagnetism from the U(1) bundle}
Starting with a U(1) bundle ${X,M,\pi,U(1)}$ over a spacetime manifold \M\ and connection one-form $\omega$ on the bundle, we introduce a local gauge connection potential, $\textbf{A}$, defined on \M\ which is a U(1) valued one form that acts on a tangent vector, $V$ to give an element of the lie algebra of U(1). It is a U(1) valued one-form, which describes how $\theta$ changes in direction $V$. All we need is a gauge potential $\textbf{A}$ defined in the single patch $U_\infty$, but it is important that a global connection $\omega$ exists and holds information about the topology of the U(1) bundle. Since U(1) is an abelian group the curvature is simply $\textbf{F}=\textbf{dA}$; and $\textbf{F}$, unlike $\textbf{A}$, is global. The first set of Maxwell's equations follows immediately from $\textbf{dd} = 0$:
\begin{equation}\label{eq:max1}
   \textbf{dF} = \textbf{ddA} = 0
\end{equation}
Where the U(1) valued curvature two form of the U(1) bundle, $\textbf{F}$, corresponds to the Faraday Tensor.

The second set of Maxwell's equations can be derived from variational principles. Using the five dimensional metric on the U(1) bundle we can construct the Ricci and scalar curvatures for the bundle and take variations of the action integral. This is the same process and principal as used to derive Einstein's equations in 4D spacetime.

The metric, $h$, on the bundle, compatible with metric, $g$, on spacetime and the connection $\omega$ is constructed as follows \cite[p135]{bleecker}: we introduce the simple metric $k^2 dx_u^2$ to measure lengths of tangent vectors on the U(1) fibres. The constant $k$ is analogous to $c$ which converts time intervals to distances and has units of $LU^{-1}$. The connection projects out the U(1) component of any vector in the bundle. We use the bundle projection $\pi$ to pull back the metric $g$ to act on vectors in the bundle.
 \begin{equation}
 h = \pi_* g + k^2 \omega \otimes \omega
 \end{equation}
 In component, terms on a patch of the manifold, we use the gauge potential $A_\mu$:
 \begin{equation}\label{eq:h}
  \textbf{h}    =
\left(%
\begin{array}{ccccc}
    g_{tt} + k^2A_t A_t & g_{xt} + k^2A_x A_t & g_{yt} + k^2A_y A_t & g_{zt} + k^2A_z A_t & k^2 A_t \\
    g_{tx} + k^2A_t A_x & g_{xx} + k^2A_x A_x & g_{yx} + k^2A_y A_x & g_{zx} + k^2A_z A_x & k^2 A_x\\
    g_{ty} + k^2A_t A_y & g_{xy} + k^2A_x A_y & g_{yy} + k^2A_y A_y & g_{zy} + k^2A_z A_y & k^2 A_y \\
    g_{tz} + k^2A_t A_z & g_{xz} + k^2A_x A_z & g_{yz} + k^2A_y A_z & g_{zz} + k^2A_z A_z & k^2 A_z \\
    k^2 A_t             & k^2 A_x             & k^2 A_y             & k^2 A_z             & k^2  \\
\end{array}%
\right)
\end{equation}
 The form of the equation is the same as the commonly used Kaluza Klein metric, but it appears canonically as the bundle metric without the ad hoc assumptions usually added in Kaluza Klein theory. $k$ is naturally a constant rather than a new field as in Kaluza Klein theory.

 Given the metric we can calculate geodesics on the bundle and construct the action:
 \begin{equation}
 S(h) = \int_X R(h) vol(h)
 \end{equation}
 where $R(h)$ is the scalar curvature of the bundle and $vol(h)$ is the volume element on the bundle evaluated using the metric $h$. Taking variations of the action with respect to the metric $h$ gives the Einstein equations in five dimensions $G_{ab} = 0$. And we can calculate the Einstein tensor in five dimensions directly from the metric \cite{williams2015}. We use Roman a,b etc for indices \{t,x,y,z,5\} and Greek $\mu \nu$ for the spacetime indices \{t,x,y,z\} :
  \begin{equation}\label{eq:max2KK}
 G_{ab} = 0 \Rightarrow G_{\mu5} = 0 \Rightarrow \frac{k^2}{2} g^{\alpha\beta} \nabla_\beta F_{\nu\alpha} = 0
 \end{equation}
 which is neatly expressed as:
 \begin{equation}\label{eq:max2}
\Rightarrow \textbf{d*F} =0
 \end{equation}
 Which is the second Maxwell equation. The spacetime components give:
 \begin{equation}\label{eq:EinsteinKK}
 G_{ab} = 0 \Rightarrow G_{\mu\nu} = -\frac{k^2}{2} g^{\alpha\beta} F_{\mu\alpha}F_{\nu\beta} + \frac{k^2}{8}g_{\mu\nu}F_{\alpha \beta}F^{\alpha\beta}
 \end{equation}
which is the usual form of the electromagnetic stress energy tensor, although it should be noted that the derivation is geometric and does not have the physical units of energy - there is no meaning to a \emph{mass}. A third equation is:
  \begin{equation}\label{eq:KKextra}
 G_{ab} = 0 \Rightarrow G_{55} = 0 \Rightarrow R = \frac{3 k^2}{4} F_{\alpha\beta}F^{\alpha\beta}
 \end{equation}
Equations \ref{eq:max2KK} and \ref{eq:EinsteinKK} are cited as the miracles of Kaluza Klein theory, although they only arise naturally from a U(1) bundle theory. Equation \ref{eq:KKextra} is enigmatic, it is unavoidable in the U(1) bundle approach and has no clear physical meaning. In the full Kaluza Klein approach $\psi = k^2$ can be treated as a scalar field rather than a physical constant as is natural in the U(1) approach. With a variable field $k$, all the equations have extra terms with derivatives of $k$, which are not shown above.

\section{Apparent charges}
Charge is normally treated as a fundamental entity with a charge density being well-defined. The total charge in a region is calculated using volume integrals. Stokes' theorem is subsequently used to relate the flux through a closed surface (e.g. $S^2$) to the actual enclosed charge. In this paper we take a different approach and use surface integrals in the region $U_\infty$ to define apparent charge. This is where our experiments take place and it is a concept of charge that is accessible and relevant. While the apparent charges are completely compatible with the existence of real charges they do not depend upon the valid application of Stokes' theorem, actual charges don't need to exist. For apparent magnetic charge:
\begin{equation}\label{eq:Qm}
Q_m = \oint_{S^2} \textbf{F}\  \overset{?}{=} \ \int_{V:S^2 = \partial V} \textbf{dF} = 0
\end{equation}
Where the second equality is valid when $V$ is a compact orientable volume (of 3 space) bounded by $S^2$. Note that $S^2$ is not necessarily the boundary of any three volume, in which case the second equality fails. One mouth of a wormhole threaded with magnetic flux is a well known counterexample. Similarly for the conserved apparent electric charge:
\begin{equation}\label{eq:Qe}
  Q_e = \oint_{S^2} \textbf{*F}  \neq   \int_{V: S^2 = \partial V} \textbf{d*F}
\end{equation}

The second equality not only requires $S^2$ to be the boundary of a compact volume, but also requires a consistent time orientation in order to have a well-defined Hodge star operator. Hence a manifold, $\M $, that is not time orientable can allow apparent electric charge from source free equations. This approach is unique in classical physics in deriving both Maxwell's equations and electric charges from a single simple assumption - the non orientability of time.

In the asymptotic patch $U_\infty$ the equations \ref{eq:max1},\ref{eq:max2},\ref{eq:Qm} and \ref{eq:Qe}, are all valid. Application of Stokes' theorem in $U_\infty$ gives the inverse square law for electric or magnetic fields from both real and apparent sources of charge.

\subsection{Conservation of apparent charge}
Conservation of charges in a space $U^3$ is conventionally proven by construction of a four volume $V^4 = U^3 \times I$ of the space and time direction \cite{eschrig2011topology}. However to use Stoke's theorem in $U_\infty$  with apparent charges. We construct a three volume $V^3 = \partial U^3_{t=0} \times I$. Conveniently, $\partial V^3 = \partial U^3_{t=0} \cup \partial U^3_{t=1}$ giving:
\begin{equation}\label{eq:con3}
0= \int_{V^3} \textbf{d*F} = \int_{\partial V^3} \textbf{*F} = - \int_{\partial U^3_{t=0}} \textbf{*F} + \int_{\partial U^3_{t=0}} \textbf{*F}
\end{equation}
The apparent charge at $t=1$ equals the apparent charge at $t=0$. The integrals all take place in $U_\infty$ in regions free of actual sources. Equation \ref{eq:con3} is equally valid for apparent magnetic charges.

Well known counterexamples to \ref{eq:Qm} and \ref{eq:Qe} exist where $V^3$ is not compact. e.g. manifolds with a defect, point or worldline removed, which are commonly used to reconcile Maxwell's equations with sources of charge. The Dirac monopole is such an example (see for example \cite{eschrig2011topology}), although the most common examples are manifolds which are not geodesically complete.

\subsection{Quantisation of apparent charges}\label{sec:charges}
The previous section that introduced charges made no use of the U(1) bundle construction or the link to orientability. However the integral of apparent magnetic charge is not only a topological invariant, but is also characteristic of the U(1) bundle. The integral is related to the first Chern cohomology class $c_1 \in H^2_{deR}(M:R)$ which is a topological characteristic of the U(1) bundle independent of the choice of connection.
\begin{equation}\label{eq:Qm3}
\frac{1}{2\pi} Q_m = \frac{1}{2\pi} \oint_{S^2} \textbf{F} = \oint_{S^2} c_1 = n \in \Z
\end{equation}
The periodicity of the U(1) bundle gives a natural metric on the fibres where a unit corresponds to one cycle. If we measure the periodicity in radians then a factor of $2 \pi$ is required - as shown above. If we use a complex phase $e^{i\phi}$ then a factor of $-i/2\pi$ is required. The magnetic charge is quantised in units of $2\pi$ provided that $H^2$ is non-trivial. For a trivial bundle $n=0$ and there cannot be magnetic charges.  In our construction the non-orientability of time requires the U(1) bundle to be non-trivial and hence $n \neq 0$ for some closed surfaces. Magnetic charges are unavoidable. Apparent electric charges can exist, but there is no topological quantisation as for magnetic charges.

Aspects of this argument are well known and usually described as a Dirac monopole \cite{eschrig2011topology}. It is worthwhile to distinguish the arguments. A key difference is the origin of the U(1) bundle: In this paper it arises naturally and inescapably from the geometry and non time orientability. In earlier works the U(1) bundle is related to the phase of the quantum mechanical wavefunction - quantum theory and some aspects of quantisation are assumed at the outset. In contrast, this paper is entirely classical and is unique in deriving both Maxwell like equations and quantised charges without invoking quantum theory. A common treatment of the Dirac monopole is to consider geodesically incomplete manifolds (by removing a point at the origin of $\R^3$), In this paper spacetime is geodesically complete and space is also compact.

Our construction \emph{requires } a non-zero magnetic charge, other approaches allow it. The status of apparent electric charges is not clear. Equation \ref{eq:max2} combined with Stokes' theorem in region $U_\infty$ ensures that there are no apparent electric charges associated with contractible surfaces $S^2$ in $U_\infty$. For surfaces enclosing the world tube Stoke's theorem breaks down. The integral theorem \ref{eq:Qm3} does not apply because $\textbf{*F}$ is not a  curvature two form of a U(1) bundle. It is not just the construction that distinguishes $\textbf{F}$, from $\textbf{*F}$, the fact that $\textbf{*F}$ is not globally defined means it cannot be a curvature form. So that electric charge is neither forbidden nor required.

To take the specific example of a time reversing wormhole described earlier. The wormhole mouths would be equal and opposite magnetic monopoles, while the structure as a whole could have net electric charge because lines of electric field flux entering one mouth would exit the other mouth with direction reversed, each mouth would have an equal charge of the \emph{same} sign. In the exterior region $U_\infty$ the structure exhibits a net electric charge and a magnetic dipole.

\subsection{The relation to electromagnetism with sources }
The constructions above derive equations of the same form as classical electromagnetism, but without the units being established, it is entirely geometric and there is no notion of mass nor units of energy. Fundamentally the equations are source-free.
However the structures we are describing with non-trivial U(1) bundles, lead to equations in $U_\infty$ that have apparent charges, potentially both electric and magnetic. The effective equations:
\begin{equation}\label{eq:max1a}
    \textbf{dF} = \textbf{*J}_m
\end{equation}
and
\begin{equation}\label{eq:max2a}
    \textbf{d*F} = \textbf{*J}_e
\end{equation}
are appropriate. Unfortunately, despite the symmetry of the equations, there is no Lagrangian that gives both \ref{eq:max1a} and \ref{eq:max2a} with source terms. (One can be constructed at the expense of introducing a second gauge potential \cite{cabibbo62}. That is a complication that we don't want to pursue here). If $F$ is a curvature two form then $\textbf{F} = \textbf{dA}$ and $\textbf{dF} = 0$ by construction, contrary to equation \ref{eq:max1a}. Therefore in $\R^4$  spacetime, we cannot construct the obvious equations for apparent magnetic charges. The symmetry of the homogeneous equations $\textbf{F}=0$ and $\textbf{*F}=0$ suggests duality transformations in $\R^4$ that can convert \textbf{E} and \textbf{B} fields (see \cite{MTW} for example), so that one or the other can have source terms but not both. However this rotation interchanges $\textbf{F}$ and $\textbf{*F}$, and in the U(1) bundle construction they are not interchangeable, only $\textbf{F}$ is a curvature two form.

The specific wormhole model described has net electric charge and no net magnetic charge. We can extend region $U_\infty$ to create a lab frame, replacing the worldtubes with point singularities. In the lab frame it is adequately described by equations \ref{eq:max1}, \ref{eq:max2} and corresponds to conventional electrodynamics in form.

\section{SI Units}
Equations \ref{eq:max1} and \ref{eq:max2} are deceptively similar to Maxwell's equations. While the mathematical form is the same, the physics content is not at all obvious. Here we comment on the tensorial forms of each entity and apply dimensional analysis to add constants where appropriate and to highlight unresolved physics issues.

Mathematically, working throughout with an orthonormal basis, it is relatively straightforward, but it becomes more complicated for physicists who want to distinguish space from time coordinates and use different units for each, the simple practice of setting $c=1$ hides many subtleties. The equations have tensorial structures like vectors and two-forms and also have coefficients that have physical dimensions giving rise to the units of measure. The numerical value of the tensor coefficients will change as the units change. Obviously adding several terms of an equation requires consistent tensor form as well as consistency of units.

We have basic length, L, and time, T, units and the metric which adds them has a conversion factor of $c$. In addition the model has a new unit, U, arising from the U(1) bundle. The connection one-form takes values in the Lie algebra of the U(1) bundle (the tangent space associated with the U(1) dimension). We denote $d\tau$ as a timelike orthonormal base vector with units $L^{-1}$ in common with $dx, dy, dz$. While $dt$ as a timelike base vector with a magnitude dependent on $c$ and having units of $T^{-1}$, $cdt =  d\tau$. For example:
\begin{equation}
A = A_0 d\tau + A_x dx + A_y dy + A_z dz = A_0 c dt + A_{x_i} dx_i
\end{equation}

where $i$ takes values $x,y,z$ In an orthonormal basis the two-forms: $F$ and $*F$ are $U$ valued with units $UL^{-2}$.  The three-forms are $\textbf{dF}$ and $\textbf{d*F}$ which are U-valued with dimension $UL^{-3}$. When working with spacetime co-ordinates the forms with a $dt$ component includes a $T^{-1}$ dimension and the coefficient has a compensating factor c with dimensions $L^{-1}T$.
%

Analysis of the units and spacetime symmetry requires a distinction between the space and the time components. Although equations \ref{eq:max1} and  \ref{eq:max2} are mathematically neat, the symmetry is more evident using the vector fields like $\textbf{E}$ and $\textbf{B}$. Here we convert to spacetime co-ordinates distinguishing time from space indices.  We write $\textbf{F}$ as:
\begin{eqnarray}\label{eq:F}
  \textbf{F}    &=& F_{0i} d\tau \wedge dx_i + F_{ij}  dx_i \wedge dx_j \\
    &=&  -E_i dt \wedge dx_i + B_k  dx_i \wedge dx_j \nonumber
\end{eqnarray}
Which defines the electric and magnetic fields with units of $U L^{-1}T^{-1}$ and $U L^{-2}$ respectively.
The dual two form $\textbf{*F}$ is more complicated:

\begin{eqnarray}\label{eq:starF}
  \textbf{*F}    &=& F_{jk} d\tau \wedge dx_i + F_{i0}  dx_j \wedge dx_k \\
    &=&  B_i c dt \wedge dx_i + E_i/c  dx_j \wedge dx_k \nonumber
\end{eqnarray}
Apparent electric charge defined by the surface integral of $*F$ (equation \ref{eq:Qe}) has units of $U$ as does magnetic charge, although $Q_m$ is the integral of the $B$ field, but $Q_e$ is the integral of $E/c$. When the time components are distinguished, Maxwell's equations \ref{eq:max1} and \ref{eq:max2} become the following four equations:
\begin{eqnarray*}
  \textbf{dF}_{ijk}    &=& 0 = \partial_i F_{jk} dx_i \wedge dx_j \wedge dx_k \\
  &\Rightarrow&  \nabla . \textbf{B} = 0 \\
  \textbf{dF}_{0jk}   &=& 0 = ( \partial_\tau F_{jk} + \partial _k F_{0j} - \partial_j F_{0k} )d\tau \wedge dx_j \wedge dx_k \\
      &=&  ( \partial_t B_i /c - \partial _k E_j/c +\partial _j E_k/c )c dt \wedge dx_j \wedge dx_k \\
  &\Rightarrow&  \partial_t \textbf{B}  + \nabla \times \textbf{E} = 0
\end{eqnarray*}
And collecting the components of $d*F = 0$:
\begin{eqnarray*}
  \textbf{d*F}_{ijk}    &=& 0 = \partial_i *F_{jk} dx_i \wedge dx_j \wedge dx_k\\
 &=&   \partial_i F_{i0} dx_i \wedge dx_j \wedge dx_k\\
   &\Rightarrow&  \nabla . \textbf{E} =0 \\
  \textbf{d*F}_{0jk}   &=& 0 = ( \partial_\tau *F_{jk} + \partial _k *F_{0j} - \partial_j *F_{0k} )d\tau \wedge dx_j \wedge dx_k\\
     &=&  ( \partial_t E_i /c^2 + \partial _k B_j -\partial _j B_k)c dt \wedge dx_j \wedge dx_k \\
  &\Rightarrow&  \partial_t  \textbf{E} /c^2  - \nabla \times \textbf{B} = 0
\end{eqnarray*}
These give the source free Maxwell equations in a form consistent with SI units. Dimensional analysis gives the following correspondence rules:
\begin{eqnarray}
  \textbf{E} &=&  \alpha c \varepsilon_0 \textbf{E}_\textrm{SI}  \\
 \textbf{B} &=&  \alpha c \frac{1}{\mu_0 c} \textbf{B}_\textrm{SI} = \alpha \varepsilon_0 \textbf{B}_\textrm{SI}
\end{eqnarray}
Where $\alpha$ is a dimensionless constant. Most other entities follow the same translation rules. The energy momentum tensor is more complicated in SI units. Einstein's equations derived from the U(1) theory \ref{eq:EinsteinKK}, is a very simple form; the left hand side is a function of spacetime curvature (units $L^{-2}$ ) and it is equated to an expression based on the U(1) curvature. A single conversion factor, $k$, is required to convert U(1) vectors to SI units of length. There is no mass dimension and the expression with the appearance of the energy momentum tensor has units of $U^2L^{-4}$.
 SI units add an additional factor $1/\mu_0$ to give units of energy density and include the factor $G/c^4$ to convert the whole RHS to units of $L^{-2}$. This gives the equivalence:
 \begin{equation}\label{Gunits}
   \frac{k^2}{2} = 8 \pi \frac{G}{c^4}. \frac{1}{\mu_0}.  \frac{1}{c^2\varepsilon_0^2}.\frac{1}{\alpha^2}
 \end{equation}

hence, the Einstein equations and curvature of spacetime provide a link to determine $k^2\alpha^2$.

\section{Equations of motion}
Although we have derived Maxwell's equations, or at least equations that are superficially the same, we do not have an equivalent of the Lorentz force law:
\begin{equation}\label{eq:Lor}
 m \ddot{\textbf{x}}= e( \textbf{E} + \textbf{v} \times \textbf{B}) = e\textbf{F.u}
\end{equation}
as written in the traditional vector form and the covariant formulation. It is an equation for the motion of a particle of charge $e$ moving with a 3-velocity $\textbf{v}$, (represented by the 4-vector $\textbf{u}$.) In fact it is sufficient to derive the Coulomb force $ e \textbf{E}$, equation \ref{eq:Lor} can then be obtained by a Lorentz transformation.

\subsection{Geodesic equations of motion}
In the spirit of our geometric U(1) bundle approach showing that a charged particle follows a geodesic path (in 5 dimensions) which corresponds to the Lorentz force law is most appealing. It does not require the notion of force, energy or mass, none of which appear in the equations naturally. to a large extent this can be done.

One of the achievements of the Kaluza-Klein theory is that the geodesics in 5D space project onto 4D spacetime as non-geodesic paths that can be interpreted as the Lorentz force law. The same is true with the U(1) construction, but the results arise naturally without the added assumptions of Kaluza Klein theories. From the metric 5D metric $h_{ab}$ \ref{eq:h} the connection can be calculated in the usual way:
\begin{equation}\label{eq:connectionfrommetric}
\Gamma^a_{bc}\equiv \frac{1}{2} h^{ad} (\partial_b h_{dc} + \partial_c h_{db} - \partial_d h_{bc} )
\end{equation}
and inserted into the geodesic equation of motion:
\begin{equation}\label{eq:KKgeodesic}
\frac{d^2 x^a}{ds^2}=-\Gamma ^a_{ bc } \frac{dx^b }{ds}\frac{dx^c}{ds}
\end{equation}
Where all indices vary over (t,x,y,z,5). This give an equation for the U component:
\begin{equation}\label{eq:KKgeodesic5}
\frac{d^2 x^5}{ds^2}=-\Gamma ^5_{ bc } \frac{dx^b }{ds}\frac{dx^c}{ds}
\end{equation}
The solution of which shows that motion in the fibre is constant (see for example \cite{wesson1995}), noting that it includes both specific motion in the $x^5$ coordinate and a contribution from the space and time vectors:
\begin{equation}\label{eq:Uconstant}
W = k^2(\frac{dx^5}{ds} + A_\alpha\frac{dx^\alpha}{ds}) = \textrm{constant}
\end{equation}
The constant of motion is derived more elegantly, but abstractly, in \cite[p144]{bleecker}. The spacetime components of \ref{eq:KKgeodesic}, with s, as proper time, give:
\begin{equation}\label{eq:KKLorentz}
\frac{d^2x^\alpha}{ds^2}=-\Gamma ^\alpha_{\mu\nu} \frac{dx^\mu }{ds}  \frac{dx^\nu}{ds} + W F^\alpha_\beta \frac{dx^\beta}{ds}
\end{equation}
This gives a familiar looking equation. The first term is the 4D geodesic motion in curved spacetime; the second is a non-geodesic term giving the lorentz force law provided that $W$ is identified with the charge to mass ratio of the test particle $W= e/m $. This is claimed to be one of the great successes of the Kaluza Klein theory, but the claim is somewhat optimistic and uncritical. The Kaluza Klein equations don't have explicit charges of electrical or magnetic monopoles and are symmetric between \textbf{E} and \textbf{B} fields. Effective charge can be measured from surface integrals, but there is no link between the notion of electric charge as a source of the fields and electric charge embodied in the constant $W$. There is no way for a mass dependency to enter the equations.

\subsection{Change of energy}
One classical route to obtain the Lorentz force law is to analyse the change in electromagnetic energy and equate it to force times distance to get the Coulomb force equation, but we have no expression for energy - although we have equation \ref{eq:EinsteinKK} with an apparent  \emph{energy momentum} term it looks like the classical term for electromagnetic energy, but without the dimensions of energy.

Another classical approach is to add matter-electromagnetism interaction terms to a Lagrangian, there are very few forms that give any non-trivial outcomes and the one we are familiar with is the simplest. We would rather not make such assumptions. After all,``no interaction with matter'' seems to be a plausible theoretical outcome. Implicit in this approach is to equate the constant $e$ in equation \ref{eq:Lor} with the charge of a particle that appears in equation \ref{eq:Qe}. The identity of the two is fundamental to classical electromagnetism, but is absent in these approaches to deriving the Lorentz force law.

\subsection{Energy conservation equations of motion}
Conservation of energy for a charge moving in a field is an alternative way to derive a force law, using $F_{em} = dU/dx$ relating the force on a charge to the change of energy. It is sufficient to derive the equations for electrostatics and then use a Lorentz transformation to get full equations. A simple example is to consider the energy change as two charged plates of a capacitor are moved closer together. With symmetry and locality assumptions, this gives $F = q.E$. Unlike the geodesic equations above, the charge in the equation is the source of the field. Essentially the field is calculated from Maxwell's equations with a source term. And because it gives a force, rather than an acceleration, a mass term is not required. And it is equally valid for apparent charges. Uniquely, this simple approach will give a version of Lorentz force law for magnetic monopoles

A more rigorous treatment is to take the divergence of the electromagnetic energy momentum tensor. However this requires equation \ref{eq:EinsteinKK} to have an additional source term. Following Misner, Thorne and Wheeler \cite[p89]{MTW}, for example, it follows that:
 \begin{equation}
 \frac{\partial T^{\mu\nu}}{\partial x_\nu} = - J_\alpha F^{\mu\alpha}
 \end{equation}
 As with the capacitor example, the charge current density is the source of the field through the equation $\textbf{d*F} = \textbf{*J}$, (the derivation also used $\textbf{dF} = 0$.) This cannot work for magnetic monopoles and $\textbf{F}$ being a curvature of the bundle, because a source term for magnetic monopoles contradicts $\textbf{dF =ddA = 0}$.

Neither approach works directly for the source free equations, which is perhaps hardly surprising: it is formulated in terms of continuous charge distributions and does not immediately apply to apparent charges. However given apparent sources, the homogeneous Maxwell equations can be solved for the regions outside the sources. In the capacitor example the results would be true for apparent electric charges in an electric field and for magnetic monopoles in a magnetic field.

All the approaches work to some extent, though all require at least one extra assumption. More positively, they all give derive the same Lorentz force law for electric charges that we are familiar with.

\section{Quantisation of electric charge}
Section \ref{sec:charges} gave a quantisation condition for magnetic charge \ref{eq:Qm3} . Although there is no equivalent topological argument for electric charges, the wormhole example gives an indirect quantisation. The wormhole has two mouths, they have equal electric charges and equal, but opposite quantised magnetic charges. For a static solution $Q_m = Q_e$. This quantisation based on the dynamics seems unconvincing because the unit of electric charge is so universal.

\section{Conclusion}
The simple and well motivated hypothesis that time is not orientable, leads to the equations of electromagnetism and the existence of electric charges. Many aspects of the paper are well known consequences of a U(1) bundle structure over a spacetime manifold, but these results give an original explanation for the U(1) bundle structure and they naturally explain the existence of charge. The work therefore extends the Kaluza Klein ideas by motivating the structure and constraints that underpin it, while also extending the homogeneous equations to include charge.

 Deriving the U(1) bundle as a measure of the twisting of the time axis is new, as is the application to manifolds that are not time orientable. Like Kaluza Klein theory, it is purely classical, but free of the complications and ad hoc assumptions required by full 5D Kaluza Klein theories. It is remarkable how so much is derived from such a simple construction without additional assumptions.

The connection with quantum theory is intriguing. Classical structures that are not time orientable have close links with quantum phenomena such as the logical structure \cite{hadley97}, particle creation and annihilation \cite{hadley2002} and spin half \cite{hadley2000} The derivation of Maxwell's equations and electric charge is just another result from the same premise of non time orientability.

Electric charges are a natural feature of the model, but quantisation only appears naturally for magnetic monopoles (which arise in pairs). However, adding elements of quantum theory to the monopole structures leads to quantisation of electric charge. The simplest such argument uses quantisation of angular momentum, but the argument fails when more than one monopole is introduced. The other approach uses a wavefunction for a charged particle \cite[p262]{eschrig2011topology}. Uniqueness of the phase of the wavefunction leads to quantisation of charge. This is a tantalising link between the complex phase of the wavefunction and the orientability of spacetime.

\begin{acknowledgements}
I am most grateful to Dr Paul Bryan for help formalising the U(1) bundle structure.
\end{acknowledgements}


%
%

\end{document}